\documentclass{epl}
\usepackage{amsmath}
\usepackage{amssymb}

\title{Gate voltage effects in capacitively coupled quantum dots}
\shorttitle{Gate voltage effects in double quantum dots.}
\author{Andrew K. Mitchell, Martin R. Galpin and David E. Logan }
\shortauthor{Andrew K. Mitchell \etal}

\institute{                    
Oxford University, Physical and Theoretical Chemistry Laboratory,\\
 South Parks Road, Oxford OX1 3QZ, UK\\
}

\newcommand{\bk}{\mathbf{k}}
\newcommand{\acre}[1]{a^\dagger_{#1}}
\newcommand{\ades}[1]{a^{\phantom\dagger}_{#1}}
\newcommand{\ccre}[1]{c^\dagger_{#1}}
\newcommand{\cdes}[1]{c^{\phantom\dagger}_{#1}}

\newcommand{\nhat}{\hat n}

\newcommand{\eref}[1]{eq.~(\ref{#1})}
\newcommand{\fref}[1]{fig.~\ref{#1}}
\newcommand{\Fref}[1]{Figure~\ref{#1}}
\newcommand{\tref}[1]{table~\ref{#1}}
\newcommand{\Tref}[1]{Table~\ref{#1}}
\newcommand{\ut}{\tilde U}
\newcommand{\met}{|\tilde\epsilon|}

\begin{document}
\maketitle

\begin{abstract}
We study a system of two symmetrical capacitively coupled quantum dots, each coupled to its own metallic lead, focusing on its evolution as a function of the gate voltage applied to each dot. Using the numerical renormalization group and poor man's scaling techniques, the low-energy Kondo scale of the model is shown to vary significantly with the gate voltage, being exponentially small when spin and pseudospin degrees of freedom dominate; but increasing to much larger values when the gate voltage is tuned close to the edges of the Coulomb blockade staircase where low-energy charge-fluctuations also enter, leading thereby to correlated electron physics on energy/temperature scales more accessible to experiment. This range of behaviour is also shown to be manifest strongly in single-particle dynamics and electron transport through each dot. 
\end{abstract}

\section{Introduction}
Recent years have seen intense investigation  of electron transport through semiconducting quantum dot devices~\cite{kouwenhoven}, leading in turn to a strong resurgence of interest in basic Kondo physics~\cite{KG}. The classic example---namely the spin-$1/2$ Kondo effect~\cite{hewson} in which the magnetic moment of an odd-electron dot is quenched by coupling to metallic leads---was observed in quantum dots almost a decade ago. Since then, a continuing goal has been to understand how \emph{coupled}, multiple quantum dot systems may lead to variants of the Kondo effect involving both spin and orbital degrees of freedom.

  In this Letter we consider probably the simplest example of such---a symmetrical double quantum dot system in which the interdot coupling is 
capacitive. Experimental realisations of such devices have appeared in the literature \cite{blick, wilhelm, chan}, and various aspects of the rich inherent physics have been uncovered in a number of theoretical papers, see e.g.~\cite{pohjola, boese, borda, lopez, ourprl, ourjpcm, mravlje}.
Here we elucidate theoretically a key underlying issue: 
the effect on transport of sweeping the dot energy levels through a wide range of values by means of a suitably applied gate voltage. 

The low-temperature transport through each dot is heavily dependent on the magnitude of the gate voltage: both the characteristic low-energy Kondo scale of the system, and the zero-bias differential conductance, vary significantly as the energy levels of the dots are lowered from the Fermi level down. If the gate voltage is tuned in such a way that only low-energy spin, or spin/orbital-pseudospin, degrees of freedom are relevant, then the Kondo scale is found (as one usually expects) to be exponentially-small. By contrast, at points where low-energy charge fluctuations can arise, we show that the Kondo scale is much larger---even when the system is intrinsically strongly correlated; leading thereby to correlated-electron physics on 
energy scales more amenable to experimental interrogation. We consider, and analyse, this behaviour within a renormalization group (RG) framework.
\section{Model and atomic limit}
The capacitively-coupled double dot model we consider consists of two correlated Anderson impurities coupled by an interdot Coulomb repulsion, with each dot 
$i=L,R$ coupled to its own separate lead; viz
\begin{equation}
\label{eq:h}
H = \sum_{i,\bk,\sigma}\epsilon_\bk\acre{\bk i \sigma}\ades{\bk i \sigma}
+ \sum_{i,\bk,\sigma}V(\acre{\bk i \sigma}\cdes{i\sigma} + \mathrm{h. c.})
+ \sum_{i}(\epsilon \nhat_i + U \nhat_{i\uparrow}\nhat_{i\downarrow})
+ U' \nhat_{L}\nhat_{R}
\end{equation}
with $\nhat_{i\sigma}=\ccre{i\sigma}\cdes{i\sigma}$ and $\nhat_i = \sum_\sigma\nhat_{i\sigma}$. The first term here describes the two metallic leads (with $i=L,R$ 
again). The second term represents the tunnel coupling of each lead to its corresponding dot, and the final terms define the energies of the double dot itself (including both intra- and interdot Coulomb repulsions $U$ and $U'$). The beauty of a typical experimental quantum dot device is that the parameters entering its effective Hamiltonian are readily adjusted. In particular the dot levels $\epsilon$ can be tuned over a wide energy range by varying the gate voltage. It is this on which we focus here, considering \eref{eq:h} as a function of $\epsilon$ when $U'<U$.
We also consider $\epsilon \le 0$ only, i.e. $\epsilon = -|\epsilon|$, since the alternative corresponds to a `frozen impurity' regime with the dots in essence unoccupied and the basic physics relatively straightforward. Moreover, the behaviour of \eref{eq:h} for $|\epsilon|$ above the particle-hole symmetric point $|\epsilon|=U' + U/2$ is obtainable from that below it by a simple particle-hole transformation; we can thus restrict our analysis to the range $0 < |\epsilon| \le U' + U/2$.

In the $V=0$ (`atomic') limit the dots are decoupled from their leads. The total dot occupation number $n$ versus $|\epsilon|$ then follows the familiar Coulomb blockade (CB) `staircase' pattern, increasing stepwise from $n=0$ to $n=4$ as $|\epsilon|$ is increased. The parameter range of interest involves only the $n=0$, $1$ and $2$ steps, the edges between them lying at $|\epsilon|=0$ ($n=0 \leftrightarrow 1$) and $|\epsilon|=U'$ ($n=1\leftrightarrow 2$). For all points along the $n=1$ step, the degenerate $(n_L,n_R)=(0,1)$ and $(1,0)$ configurations lie lowest in energy, whereas on the $n=2$ step the ground state is $(1,1)$; and at the $n=1\leftrightarrow 2$ CB edge these states are all degenerate.

In what follows we elucidate the behaviour of \eref{eq:h} with increasing 
$|\epsilon|$ when the interacting dots are connected to their leads. We show that for all $U'<U$, the `atomic limit' degrees of freedom described above are always quenched as $T\to 0$, such that the ground state is a non-degenerate Fermi liquid phase throughout\footnote{It is known  \cite{ourprl, ourjpcm} that in the $n=2$ regime for a critical $U'_\mathrm{c}>U$, the model undergoes a quantum phase transition to a non-Fermi-liquid, doubly-degenerate `charge ordered' phase. }
More central, however, is how the \emph{nature} of this quenching depends on the gate voltage $|\epsilon|$. Near the midpoints of the CB steps, the dot degrees of freedom (be they spin or a combination of spin and orbital-pseudospin) are Kondo-quenched on exponentially small temperature scales. Near the edges of the steps however, where charge fluctuations are important, the behaviour is markedly different. The Kondo temperature $T_K$ is found to be \emph{much} larger, on the order of the dot-lead hybridisation parameter $\Gamma = \pi\rho V^{2}$ (with $\rho$ the host/lead density of states). Since $T_K$ sets the natural scale below which the many-body physics arises, its enhancement means that non-trivial coherent electron transport occurs over a much larger temperature window.
\section{NRG and fixed points}
Results discussed in the next section have been obtained using Wilson's powerful
numerical renormalization group (NRG) approach \cite{kww1,kww2}, technical discussion of how to apply which to the Hamiltonian \eref{eq:h} can be found in \cite{ourjpcm}. NRG is of course an iterative procedure that converges ultimately to a stable \emph{fixed point}\cite{kww1,kww2}, the nature of which indicates the structure of the ground state; and the NRG flows that lead to it (via the unstable fixed points) can be used to obtain static and dynamic properties on all important temperature and energy scales respectively.

Our results will be interpreted in terms of six distinct fixed points, the Hamiltonians for which may be obtained simply by setting each of the bare parameters of \eref{eq:h} to either $0$ or $\infty$. The simplest is the unstable free-orbital (FO) fixed point\cite{kww1}, where $\epsilon = U = U' = \Gamma = 0$. Four more unstable fixed points arise from $\Gamma = 0$ and $U=\infty$ (with fixed ratios of $\epsilon/U$ and $U'/U$), each corresponding to one of the atomic limit states discussed above. Finally there is the stable strong-coupling (SC) fixed point in which $\Gamma = \infty$, to which all NRG flows ultimately tend. \Tref{tab:fp} lists all these fixed points, the conditions under which they obtain, the allowed dot configurations and their degeneracies.
\begin{table}
\caption{\label{tab:fp}NRG fixed points discussed in this work. For detailed discussion, see text.}
\begin{center}
\begin{tabular}{lccccll}
Label & $\Gamma$ & $U$ &  $U'$ & $|\epsilon|$ & Allowed dot configurations& Degeneracy\\
\hline
FO & 0 & 0 & 0 & 0 & All & 16\\
VF$_{01}$ & 0 & $\infty$ & $U/2$  & 0 & (0,0), (0,1), (1,0) & 5\\
LM$^{SU(4)}_{1}$ & 0 & $\infty$ & $U/2$  & $U/4$ &(0,1), (1,0)& 4\\
VF$_{12}$ & 0 & $\infty$ & $U/2$  & $U/2$ & (0,1), (1,0), (1,1) &8\\
LM$^{SU(2)}_{2}$ & 0 & $\infty$ & $0$  & $U/2$ & (1,1) & 4\\
SC & $\infty$ & $0$ & $0$  & $0$ & N/A (since $\Gamma = \infty$) & 1
\end{tabular}
\end{center}
\end{table}

As seen in \tref{tab:fp}, the unstable fixed points divide into two sets. The first comprises the `local moment' (LM) fixed points, in which the total dot charge is fixed and hence the only possible dot degrees-of-freedom are those of spin and orbital pseudospin. The second set contains the `valence fluctuation' (VF) fixed points, in which charge fluctuations of the double dot also enter. It is the existence of these two distinct types of fixed point that leads to 
interesting physics, as now explained. With decreasing temperature/energy scale, all RG flows begin close to the FO fixed point (we take the host bandwidth $D$ to be the largest energy scale of the model). But for fixed $U$ and $U'$, the subsequent flows then depend solely on the magnitude of $|\epsilon|$ (as controlled by a gate voltage); such that the flow from the final unstable fixed point to the stable SC fixed point determines the characteristic low-energy scale.

For most values of $|\epsilon|$---sufficiently far from the CB edges---the NRG flows are found to pass essentially from FO to one of the LM fixed points, before crossing over to the stable SC fixed point where the dot local moments are quenched. The effective low-energy models here are thus Kondoesque and the low-energy Kondo scale $T_K$ on which the SC fixed point is reached is exponentially-small in $U/\Gamma$ and $U'/\Gamma$ (as considered further below).

If however $|\epsilon|$ is instead tuned to within $\mathcal{O}(\Gamma)$ of a CB edge, the VF fixed points also come into play and the resulting physics is quite different. This is most easily seen at one of the CB edges itself, where there is no sign of any LM fixed points at all. The RG flow instead goes from FO to SC via VF, and the Kondo scale is found to be a simple multiple of $\Gamma$, virtually independent of the bare $U$ and $U'$. So by tuning the system from one CB step to another, one can produce dramatic changes in its Kondo scale and low-energy degrees-of-freedom, which are of course evident in both statics and dynamics.

Before discussing numerical results, we make one further remark. The Kondo physics---and in particular the exponentially small $T_K$---associated with the LM $\to$ SC crossover is a well-known manifestation of strong electron correlation. But while the low-energy scales associated with the CB edges depend only on the bare energy scale $\Gamma$, the corresponding electron dynamics are nonetheless still correlated. Indeed for $U/\Gamma$ sufficiently large, the only region in which electron correlation is \emph{not} important is $\epsilon \gtrsim \Gamma$. As we shall see below, this means that the physics associated with crossing the $n=1\leftrightarrow2$ CB edge is distinct from that arising near $0\leftrightarrow1$ \cite{pohjola}: only the former involves correlated physics throughout, and it is this to which we direct the majority of our attention.

\section{Results}
We first present some static properties, beginning with the entropy. Since each of the fixed points in \tref{tab:fp} has a well defined dot degeneracy, the contribution to the entropy of the system from the dots alone, denoted by $S_\mathrm{imp}$, can be used to determine how the NRG flows depend on the bare parameters. In what follows, it is convenient to work with the dimensionless parameters $\met=|\epsilon|/\pi\Gamma$, $\ut=U/\pi\Gamma$ and $\ut'=U'/\pi\Gamma$.

\begin{figure}[t]
\caption{\label{fig:fig1} For $(\ut, \ut')=(14,12)$, (a) entropy, 
$S_\mathrm{imp}$ \emph{vs} $T/\Gamma$ 
for $\met=\ut'/2=6$ (solid line), $\met=11$ (short dashed), $\met=\ut'=12$ (long dashed) and $\met=\ut'+\ut/2=19$ (dot-dashed). (b) $T=0$
charge susceptibility $D\chi^+_{c,\mathrm{imp}}(0)$ \emph{vs} $\met$.}
\begin{center}
\includegraphics{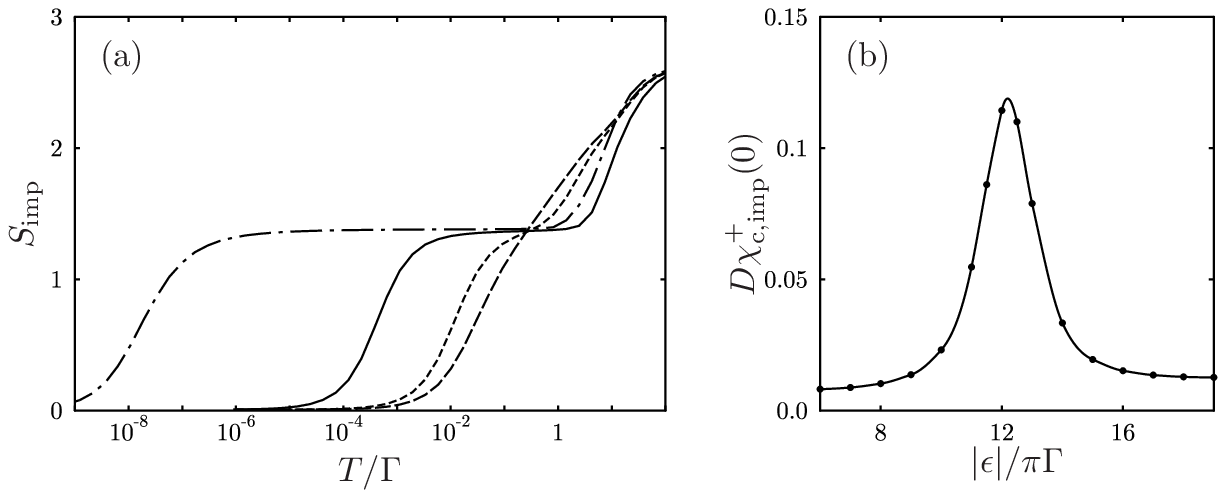}
\end{center}
\end{figure}

 \Fref{fig:fig1}(a) shows $S_\mathrm{imp}$ as a function of temperature for fixed $(\ut,\ut') = (14,12)$ at four representative points along the $\met$ line. These are the three points of `symmetry' at $\met = \ut'/2$ (centre of $n=1$ CB step, solid line), $\met =\ut'$ ($n=1\leftrightarrow2$ CB edge, long dashes) and 
$\met =\ut'+\ut/2$ (centre of $n=2$ CB step, dot-dashed); plus a fourth point at 
$\met = 11$ (short dashes) which illustrates the behaviour slightly away from the 
$n=1\leftrightarrow2$ CB edge. In all cases, the high-temperature $\ln 16$ 
free-orbital behaviour is seen to cross over ultimately to the singlet 
strong-coupling regime below some characteristic low temperature scale, but at intermediate temperatures the physics is clearly different in each case. 

The $\met = \ut'/2$ and $\ut'+\ut/2$ curves, corresponding to the centres of the $n=1$ and $n=2$ CB steps respectively, show the generic Kondoesque behaviour described  above. The long intermediate plateaus of $\ln 4$ each indicate a flow that passes very close to one of the two LM fixed points in \tref{tab:fp}. Specifically, for $\met = \ut'/2$ the fixed point is LM$_1^{SU(4)}$ and the low-energy effective model is the one-electron $SU(4)$ spin/orbital Kondo model \cite{borda, boese}. For $\met = \ut'+\ut/2$ by contrast, it is the LM$_2^{SU(2)}$ fixed point that is approached, and at low energies one observes an effective pair of uncoupled $SU(2)$ spin-Kondo models\cite{ourprl}. 

The $\met=\ut'$ entropy curve in \fref{fig:fig1}(a) illustrates the distinct behaviour that arises when the RG flow passes close to a VF fixed point---in this case the VF$_{12}$ fixed point of \tref{tab:fp}. No local moment plateaus are observed; low-energy charge fluctuations are now possible and the entropy instead drops to zero on a much higher temperature scale characterised by the hybridisation parameter $\Gamma$. When one is sufficiently close to, but not at, the CB edge at  $\met=\ut'$, both VF and LM fixed points are observed, as in the $\met = 11$ ($= \ut'- 1$) curve in \fref{fig:fig1}(a); here the entropy follows the $\met=\ut'$ form at relatively high temperatures before crossing over to a characteristic $\ln 4$ LM shoulder and then finally zero as $T\to 0$.

To illustrate charge fluctuations near the $1\leftrightarrow2$ CB edge, \fref{fig:fig1}(b) shows the $T=0$ charge susceptibility $D\chi_{c,\mathrm{imp}}^+(0)$ \emph{vs} $\met$ (with $4T\chi_{c}^{+}(T)=\langle[\hat{N}-\langle\hat{N}\rangle]^2\rangle$ and $\hat{N}$ the total charge operator \cite{ourjpcm}). A significant peak occurs at the CB edge at $\met=\ut' = 12$, reflecting the availability here of the low-energy charge degrees-of-freedom. Away from this point however, the low-energy effective model involves only spin or spin/orbital degrees of freedom
and $\chi_{c,\mathrm{imp}}^+(0)$ is very small (vanishing as $\ut\to\infty$).
The HWHM of the peak gives an estimate of the range over which low-energy charge fluctuations are significant: here we see it is $\mathcal O(1)$, and hence the charge degrees-of-freedom are important only when $|\epsilon|$ is within 
$\mathcal O(\Gamma)$ of the CB edge. 

\begin{figure}[t]
\caption{\label{fig:entropy}
Variation of $T_K$ with $\met$. 
(a) $T_K/\Gamma$ \emph{vs} $\met$ for $(\ut,\ut')=(14,12)$: solid line from NRG,  dashed lines from poor man's scaling.
The dotted line shows the average total dot occupation number $n$ (right-hand scale). (b) NRG $T_K$ versus $\met$ for different inter- and intradot coupling strengths: $(\ut,\ut')=(10,8)$ (solid line),  $(12,10)$ (long dashed), $(14,12)$ (short dashed) and $(18,12)$ (dot dashed).}
\begin{center}
\label{fig:fig2}
\includegraphics{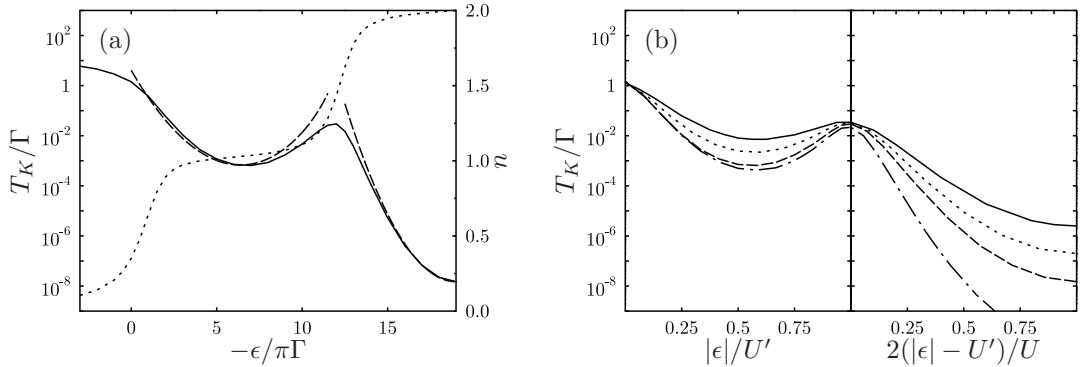}
\end{center}
\end{figure}

We now analyse the Kondo temperature $T_K$ as a function of $\met$ (with $T_K$ taken to be the temperature at which the impurity spin susceptibility satisfies $k_B T\chi_{s,\mathrm{imp}}(T)/(g\mu_B)^2 = 0.1$). As seen already in \fref{fig:fig1}(a), $T_K$ varies by many orders of magnitude as $\epsilon$ is lowered: this is shown more clearly in \fref{fig:fig2}(a) where we plot $\ln T_K/D$ versus $-\epsilon/\pi\Gamma$ for the same fixed $(\ut, \ut')=(14,12)$ (solid line).  Also shown is the total dot charge $n$ (dotted line), showing the characteristic `rounding' of the CB staircase on coupling to the leads. It is quite clear that $T_K$ is a minimum when $n\simeq 1$ or $2$, and a maximum of $\mathcal{O}(\Gamma)$ at the $1\leftrightarrow 2$ CB edge when $n$ can fluctuate easily between the two. Note that $T_K$ also increases as one moves from the $n\simeq 1$ $SU(4)$ Kondo regime towards the $0\leftrightarrow1$ CB edge; but in contrast to the behaviour just described continues to increase as $\epsilon$ becomes positive, reflecting the essential absence of electron correlation in the $n\simeq 0$ regime and a Kondo scale naturally on the order of 
$\Gamma$.

The dependence of $T_K$ on $\met$ can be understood more precisely from the effective Kondo models in the $n=1$ and $n=2$ regimes, obtained by straightforward Schrieffer-Wolff transformations and then analysed using Anderson's ``poor man's'' scaling\cite{hewson}. We find in both cases that the leading exponential form of $T_K/D$ is $\exp[-(2\rho J)^{-1}]$, with $\rho J$ given by
\begin{equation}
\label{eq:rhoj}
\pi^2\rho J \sim 
\begin{cases} 
2/\met-(\met-\ut')^{-1} - (\met-\ut)^{-1}& \mbox{for }\met\sim\ut'/2 \\
({\met-\ut'})^{-1} - ({\met-\ut'-\ut})^{-1} & \mbox{for }\met\sim\ut'+\ut/2 
\end{cases}
\end{equation}
in the strongly correlated ($U\gg\Gamma$) limit. These results for $T_K$ are plotted as dashed lines in \fref{fig:fig2}(a), for the same fixed $\ut$ and $\ut'$, each scaled by a single multiplicative factor onto the NRG result at the appropriate minimum. The strong agreement with numerical and analytical results over a wide range of $\met$,
provides further confirmation that the low-energy physics of the dots is essentially captured by Kondo models of fixed $n$, except when $\met$ is within $\mathcal{O}(\Gamma)$ of a CB edges and charge fluctuations become important.

In \Fref{fig:fig2}(b) we also plot $\ln T_K/D$ versus $\met$ for a range of different coupling strengths. The key point here is that while the scales in the Kondo regimes clearly vary significantly with all the bare parameters of the model (as seen from \eref{eq:rhoj}), those associated with the VF points are largely dependent on the hybridisation strength $\Gamma$ only. For this reason, one would expect all the low-temperature thermodynamics of the model to be essentially independent of $\ut$ and $\ut'$ near the CB edges, and we have indeed found this to be true. 

We now turn to dynamics, focusing on the local single-particle spectrum 
$D_i(\omega)$ defined in terms of the 
retarded Green function $G_i(t)=-i\theta(t)\langle\{\cdes{i\sigma}(t),\ccre{i\sigma}\}\rangle$ via $D_i(\omega)=-\mathrm{Im}G_i(\omega)$, since this quantity probes the charge fluctuations that control transport across a given dot $i$ (the model considered, where each dot hybridises to its respective lead, is formally equivalent to a 4-lead device where each dot hybridises to two leads with strength $\Gamma/2$).
As shown in \cite{meir, ourjpcm}, the $T=0$ zero-bias differential conductance ($G$) of each dot is related exactly to the 
zero-frequency spectrum by $G/(2e^2/h) = \pi\Gamma D_i(0)$
(and the spectrum at non-zero $\omega$ provides an approximation to the conductance at source-drain bias $V=\omega/e$).

\begin{figure}[t]
\caption{\label{fig:fig3}Spectral density of dot $i$ for fixed $(\ut,\ut')=(10,8)$. (a) and (b) show the evolution of $\pi\Gamma D_i(\omega)$ with increasing $\met$: specifically, $\met= 4$, $6$ and $7$ and $8$  $(=\ut')$ [(a): dot-dashed, dotted, dashed and solid respectively]; and $10$, $11$ and $13$ [(b): dashed, dotted and solid]. (c) $T=0$
zero-bias conductance $G/(2e^2/h)=\pi\Gamma D_{i}(0)$, as a function of $\met$.}
\begin{center}
\includegraphics{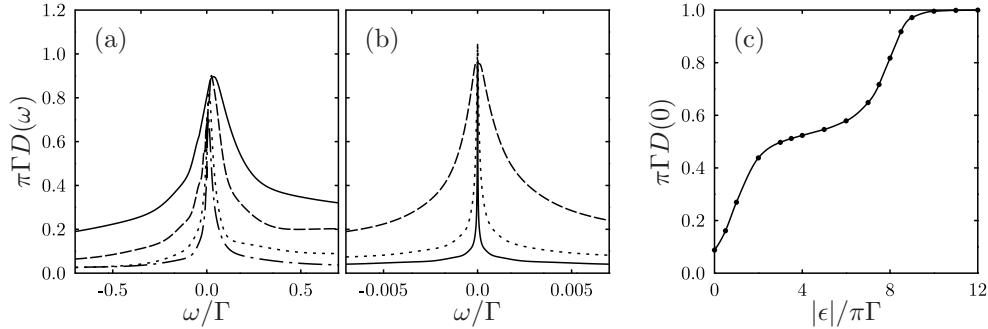}
\end{center}
\end{figure}

The NRG technique can be used in practice to calculate $D_i(\omega)$ on all frequency scales of interest\cite{CHZ}. Near the centre of each CB step, we find that the high-frequency spectrum is dominated by its Hubbard satellites, corresponding physically to sequential tunnelling through the dot. Since these features are readily explained simply as broadened versions of the poles arising in the atomic limit $V=0$, we do not discuss them further here.  More interesting is the low-frequency form of the spectrum, since this corresponds to coherent tunnelling through the dot mediated by the Kondo effect. In figures \ref{fig:fig3}(a),(b), we illustrate the evolution of the low-energy Kondo resonance with increasing $\met$ for fixed $(\ut,\ut')=(10,8)$.  \Fref{fig:fig3}(a) shows the spectrum for $\met= 4$, $6$, $7$ and $8$ (dot-dashed, dotted, dashed and solid respectively), and \fref{fig:fig3}(b) shows $\met=10$, $11$ and $13$ (dashed, dotted and solid). Note the very different frequency scales in each case. The width of the resonance, being proportional to $T_K$, is seen to first increase with $\met$ as one moves from the centre of the $n=1$ CB step ($\met=\ut'/2=4$) to the maximum at the $1\leftrightarrow2$ CB edge $\met=\ut'=8$, where $T_K\sim {\cal{O}}(\Gamma)$.
Thereafter, the resonance narrows until the particle-hole symmetric point at $\met=\ut'+\ut/2=13$ where $T_K$ is a minimum. In addition to the changing width of the resonance with $\met$, there is also a definite increase in asymmetry near the $1\leftrightarrow2$ CB edge, since here the addition and removal of low-energy electrons to and from the ground state clearly occur with different weights.

Finally, the zero-frequency spectral density $\pi\Gamma D_{i}(0)$, i.e. the 
zero-bias differential conductance, is shown as a function of $\met$ in \fref{fig:fig3}(c). Its stepwise increase is readily explained by Fermi-liquid theory, as one can show that $\pi\Gamma D_i(0) = \sin^2(\pi n_i/2)$ with $n_i$ ($\equiv n/2$) the occupation number of dot $i$, whence $\pi\Gamma D_i(0)$ itself mimics the characteristic CB staircase form. For $T \ll T_K$, the zero-bias differential conductance provides as such a direct means of measuring the average occupation number of each dot. 

\section{Conclusion}
The capacitively coupled double dot system described by \eref{eq:h} shows a rich range of behaviour as the energy levels of the dots are varied, which we have 
elucidated by a combination of NRG and poor man's scaling techniques.  By applying a suitable gate voltage, the system can be tuned to display either an $SU(2)\times SU(2)$ spin-Kondo effect ($n=2$), an $SU(4)$ spin/orbital Kondo effect ($n=1$), or correlated mixed-valence physics where the Kondo scale is substantially enhanced and non-trivial coherent electron transport thus arises over a much wider range of temperatures. It can moreover be shown that this essential underlying physics is robust to inclusion of a direct interdot hopping (`$t$'), provided $|t| \lesssim T_{K}$.

\acknowledgments
We are grateful to the EPSRC for financial support.

\end{document}